\documentclass[nofootinbib,superscriptaddress,twocolumn,showpacs,preprintnumbers,amsmath,amssymb]{revtex4-1}

\usepackage{graphicx}
\usepackage{grffile}
\usepackage{slashed}
\usepackage{bbm}
\usepackage{footmisc}
\usepackage{multirow}

\usepackage{hyperref}
\usepackage{amssymb}
\usepackage{amsmath}
\usepackage{color}
\usepackage{xspace}
\usepackage{caption}
\usepackage{subcaption}
\usepackage{natbib}

\usepackage{datetime}
\usepackage{color,fancybox}

\newcommand{\GeV}{\ensuremath{\,\mathrm{GeV}}\xspace}
\newcommand{\TeV}{\ensuremath{\,\mathrm{TeV}}\xspace}

\newcommand{\fb}{\ensuremath{\,\mathrm{fb}}\xspace}

\newcommand{\bea}{\begin{eqnarray}}
\newcommand{\eea}{\end{eqnarray}}

\newcommand{\fig}[1]{Fig.~\ref{#1}}

\newcommand{\ZAjj}{\ensuremath{Z_\ell \gamma jj}\xspace}

\newcommand{\znajj}{\ensuremath{Z_\nu \gamma jj}\xspace}
\newcommand{\ZnAjj}{\znajj}
\newcommand{\ZZjjl}{\ensuremath{pp \to \ell_1^+ \ell_1^- \, \ell_2^+ \ell_2^-~jj ~+~X}\xspace}
\newcommand{\ZlZnjj}{\ensuremath{pp \to \ell^+\ell^- \nu \, \bar{\nu}~jj ~+~X}\xspace}

\newcommand{\ZZjj}{\ensuremath{Z_\ell Z_\ell jj}\xspace}

\newcommand{\HTstar}{\ensuremath{H_{T^*}}}

\begin{document}
\title{$Z\gamma$ production in vector-boson scattering at next-to-leading
  order QCD}

\preprint{IFIC/17-40\;\;FT-UV-17-2308\;\;KA-TP-15-2017\;\;MPP-2017-46}
\author{Francisco~Campanario}
\email{francisco.campanario@ific.uv.es}
\affiliation{Theory Division, IFIC, University of Valencia-CSIC, E-46980
  Paterna, Valencia, Spain}
\author{Matthias~Kerner}
\email{kerner@mpp.mpg.de}
\affiliation{Max Planck Institute for Physics, F\"ohringer Ring 6, D-80805
M\"unchen, Germany}
\author{Dieter Zeppenfeld}
\email{dieter.zeppenfeld@kit.edu}
\affiliation{Institute for Theoretical Physics, KIT, 76128 Karlsruhe, Germany}

\begin{abstract}
Cross sections and differential distributions for $Z\gamma$ production
in association with two jets via vector boson fusion are presented at
next-to-leading order in QCD. The leptonic decays of the $Z$ boson
with full off-shell effects and spin correlations are taken into
account. The uncertainties due to different scale choices and pdf sets
are studied. Furthermore, we analyze the effect of including anomalous
quartic gauge couplings at NLO QCD.
\end{abstract}


\maketitle

\section{Introduction}
\label{sec:intro}
%
%
%
%
Electroweak photon production in association with two charged leptons
and two jets is an important channel at the LHC since it provides
information on weak boson scattering. It is also sensitive to beyond
standard model (BSM) physics via anomalous gauge boson couplings.

At LO, there are two mechanism to produce $\ell^+\ell^- \gamma jj$
events at the LHC. The QCD-induced of order ${\cal O}(\alpha_s^2
\alpha^3)$ and the EW-induced of order ${\cal O}(\alpha^5)$ which is
further classified into the s-channel contributions, given mainly by
tri-boson production with a subsequent hadronic decay of one of the
vector bosons, and the t/u-channel vector boson fusion~(VBF)
contributions.

The QCD-induced mechanism is considered to be an irreducible
background of the EW mechanism due to the lack of weak boson
scattering and quartic gauge boson couplings, $VV \to VV$. Despite the
$\alpha_s^2/\alpha^2$ enhancement, its contribution is comparable in
typical VBF searches.
%
The interference effects among the different mechanism/channels are
expected to be small in LHC measurements. A dedicated study of these
effects was carried out in Ref.~\cite{Campanario:2013gea} for same sign
$WWjj$ production, where interference effects are expected to be larger
due to the absence of gluon initiated processes and the fixed
chirality of the quark lines.

The NLO QCD corrections for the QCD-induced mechanism were given in
Ref.~\cite{Campanario:2014wga}. The corrections are moderate, if adequate
central scales are chosen, but phase space dependent and lead to a
significant reduction of the scale uncertainty.

The NLO QCD corrections of the tri-boson production processes were
first computed including the leptonic decays in Refs.~\cite{Bozzi:2009ig,Bozzi:2010sj,Bozzi:2011en} and
afterwards including the hadronic decays in Ref.~\cite{Feigl:2013naa,Baglio:2014uba}. They
turned out to be large, around 70$\%$, and not covered by the scale
uncertainties. This is due to logarithmically enhanced
configurations~\cite{Campanario:2013wta,Baglio:2013toa} and new channels opening up at
NLO.

Both the QCD-induced and tri-boson production processes are available
in the VBFNLO
package~\cite{Arnold:2008rz,Baglio:2014uba}.

\begin{figure}[ht!]
\centering
\includegraphics[width=0.49\textwidth]{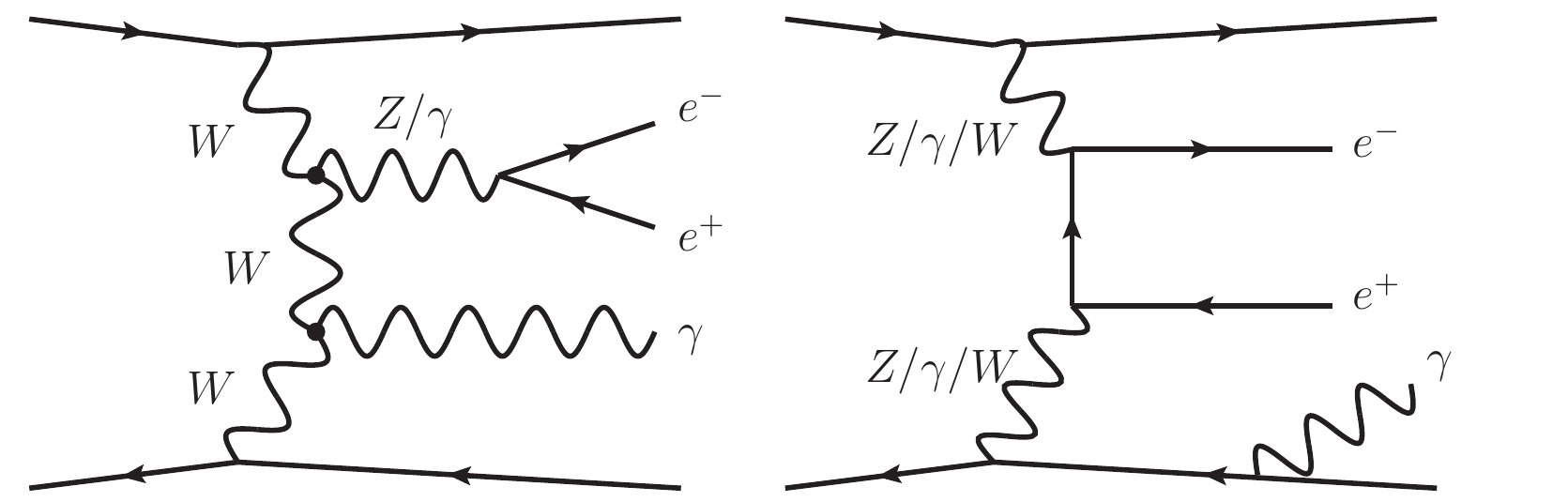}
\caption{Representative Feynman diagrams at LO.
}
\label{fig:FeynmanDiagrams}
\end{figure}

In this paper, we provide results at NLO QCD for the VBF t/u-channel
for the processes
\begin{eqnarray} 
pp &\to &\ell^+ \ell^- \gamma jj + X, \quad "Z_\ell \gamma jj",
\\\label{processE} pp &\to &\nu \bar{\nu} \gamma jj + X, \quad "Z_\nu
\gamma jj",
\end{eqnarray}
focusing mainly on the charged leptonic channel. Representative Feynman 
diagrams are shown in Fig.~\ref{fig:FeynmanDiagrams}.
The bulk of the cross section for both processes comes from regions of
the phase space where the intermediate $Z$ boson is approximately
on-shell.
For simplicity, in the following, we often refer to the processes by
"$Z_\ell \gamma jj$'' or "$Z_\nu \gamma jj$'' production, although we
will consider all off-shell effects, non-resonant diagrams and spin
correlations.
%

Furthermore, using an Effective Theory (EFT) approach, we will study,
at NLO QCD, BSM effects due to anomalous quartic gauge couplings.

The processes considered here
have been implemented in
VBFNLO~\cite{Arnold:2008rz,Baglio:2014uba}, a parton
level Monte Carlo program which allows the definition of general
acceptance cuts and distributions.

The rest of the paper is organized as follows: in
section~\ref{sec:calc} the calculational setup as well as the checks
performed to ensure the correctness of the calculation are given. In
section~\ref{sec:results}, results at the integrated cross section
level, for differential distributions, as well as for anomalous
couplings studies are presented. Finally, we present our conclusions
in section~\ref{sec:concl}.
 
\section{Calculational Setup}
\label{sec:calc}
%
%
%
%
The calculation method of the "$Z_\ell \gamma jj$" and "$Z_\nu \gamma
jj$" production follows closely the one of \ZZjjl~\cite{Jager:2006cp} (called
from now on \ZZjj production for simplicity) and other VBF channels
implemented in VBFNLO.

We work in the VBF approximation and, therefore, only the t/u-channel
Feynman diagrams, neglecting the interference between them, are
considered. The s-channel contributions at NLO are accessible in
VBFNLO via ``$ZAV$'' production, with $V\in (W,Z,\gamma)$ decaying
hadronically. Their size as well as the interference effects are small
once typical VBF cuts are applied.

We work in the five flavor scheme. For the potentially resonating
massive vector bosons, we use a modified complex mass scheme as
implemented in MadGraph~\cite{Alwall:2007st}.

Technically, to obtain the NLO QCD corrections to EW t/u-channel
contributions of \ZAjj and \ZnAjj~production, we adapt the code with
some modifications from the process \ZZjjl and \ZlZnjj implemented in
VBFNLO, respectively. Here, to be self-contained, 
we give a brief description of the method
used  for the \ZAjj process. The \ZnAjj channel is
calculated analogously.

To compute the amplitudes, we use the helicity 
formalism of Ref.~\cite{Hagiwara:1988pp}. This allows us to factorize 
amplitudes into a QCD part and electroweak factors.
The latter contain not only the decay currents for 
$V \to \ell^+\ell^-$ and $\tilde{V} \to \ell^+\ell^- \gamma$ with
$V, \tilde{V} \in (Z,\gamma^*)$, but also "leptonic tensors" for 
$VV/W^+W^- \to \ell^+\ell^-/\gamma/\ell^+\ell^-\gamma$, containing the scattering
of the $t$-channel vector bosons connecting the quark lines.
The EW currents and tensors are calculated only once per phase space
point using the routines of the HELAS package~\cite{Murayama:1992gi}. Afterwards,
the full LO amplitudes for all sub-processes, $q_1 q_2 \to q_3 q_4
\ell^+\ell^-\gamma$ and crossing related ones, can be easily obtained using the pre-calculated
structures. Similarly, we obtain the real emission amplitudes by
adding an additional gluon emission to a quark line.


For the virtual corrections, we do not consider graphs with a gluon exchange
between the two quark lines. Due to the color structure of the
amplitude, these would only give non-vanishing contributions for the
interferences of $t$- and $u$-channel diagrams, which are phase-space
suppresses and neglected in the VBF approximation.
Thus only corrections to a quark line (the upper or the lower one)
with up to three bosons emitted have to be considered. We make use of
the routines Boxline and Penline computed in Ref.~\cite{Campanario:2011cs}, which
respectively combine all the loop diagrams to a quark line with two or
three bosons emitted in a fixed order permutation of the external
bosons. The full amplitude is obtained by including all physically
allowed permutations.

We use the Catani-Seymour formalism~\cite{Catani:1996pk} to deal with the
infrared divergences and we follow Ref.~\cite{Figy:2003nv} to obtain individual
factorization and renormalization scales for each of the two quark lines. This is possible because
they each form
a gauge invariant sub-set.

To deal with the real photon in the final state, we implement the
Frixione smooth cone isolation cut~\cite{Frixione:1998jh}, which preserves IR safety, without
the need of introducing photon fragmentation functions. Additionally,
we split the phase space integration into two separated regions to improve the
convergence of the Monte Carlo integration. 
The regions are generated as double EW boson production with
(approximately) on-shell $Z\to \ell\ell $ decay or as Z production 
with $Z\to \ell\ell \gamma $ three-body
decay, respectively, and they are chosen according to whether $m(\ell
\ell \gamma) $ or $m(\ell \ell ) $ is closer to $M_Z$.
The final
result is obtained by adding the two integrals. The presence of
intermediate off-shell photons, which are far from on-shell for
typical lepton cuts, does not pose numerical problems and their
contribution is integrated with the $Z$ ones. For the $Z_\nu \gamma
jj$ production channel, this phase space splitting is not necessary.

To ensure the correctness of our results, we have cross-checked our LO
and real matrix elements with Madgraph~\cite{Alwall:2007st} and we
compared the integrated cross sections with Sherpa, finding agreement
at the machine precision and at the per mille level, respectively. For
the real emission contributions, we need to subtract the s-channel
contributions explicitly which are included in Sherpa, otherwise,
agreement at the few percent level is found for typical VBF
cuts. Thus, the neglected s-channel contributions only give a
noticeable contribution to the real emission, leading to an increase
at the few per mille level of the total NLO QCD cross section. Hence,
they can be safely neglected. In the following, we discard these
contributions. However, they can be included easily within the VBFNLO
package at NLO QCD, by adding the cross sections for triple electroweak boson 
production. 

Furthermore, we have checked the convergence of the Catani-Seymour
subtraction and, for the virtual contributions, the factorization of
the poles, as well as gauge and parametrization
invariance~\cite{Campanario:2011cs}.

The numerical stability of the calculation of the virtual amplitudes
is controlled with the use of Ward
identities~\cite{Campanario:2011cs}.  The amplitude is set to zero if
they are not satisfied with a precision better than one per mille. The fraction
of phase space points rejected is around the per mille level and thus,
the error induced by this procedure is negligible since the total virtual 
contributions, after analytic cancellation of infrared divergences, are at the level of a few per cent.
%
%
%
%
\section{Phenomenological results}
\label{sec:results}
%
%
%
%
%
%
%
\begin{figure*}[ht!]
\centering
\includegraphics[width=0.49\textwidth]{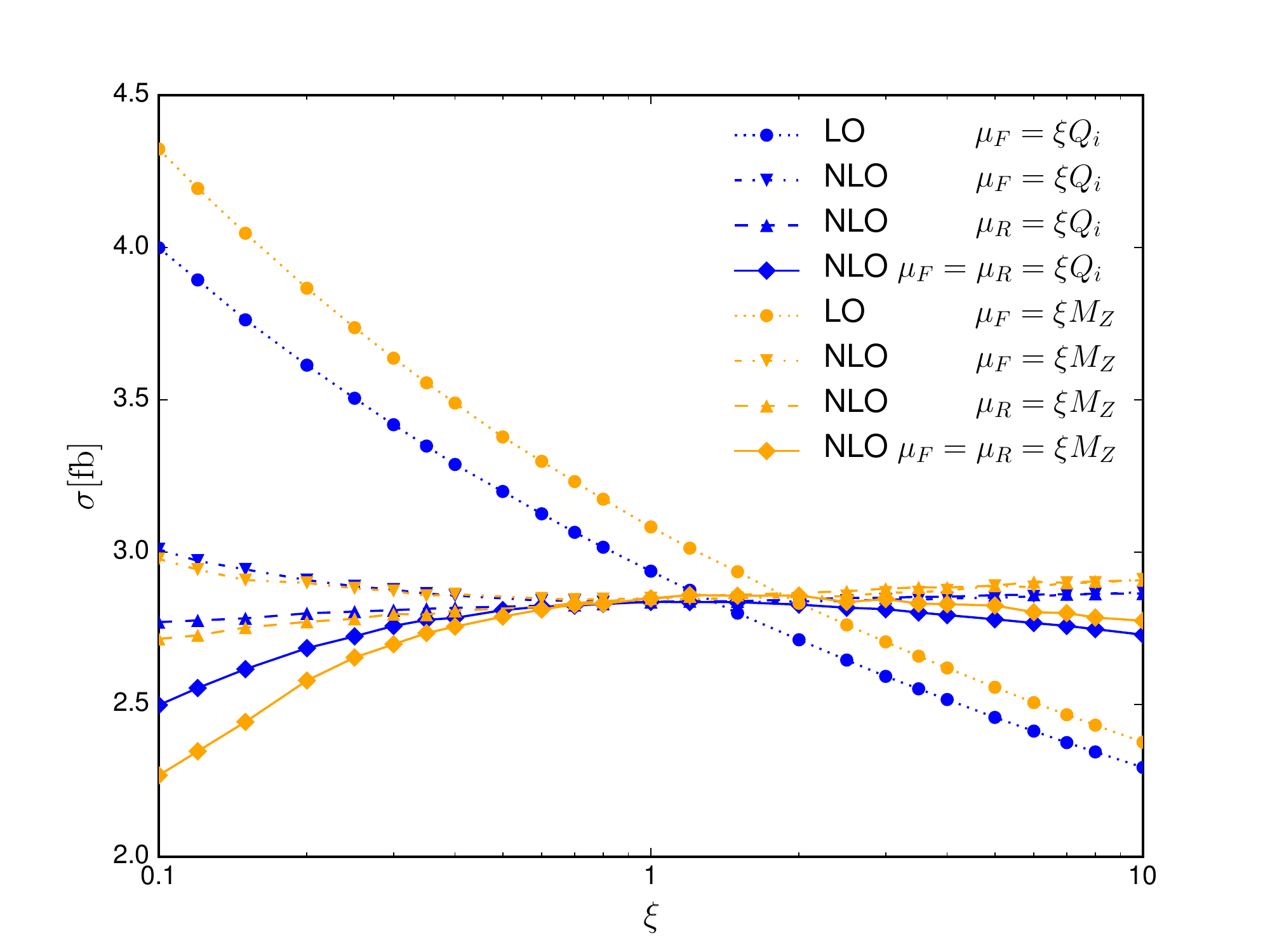}
\hfill
\includegraphics[width=0.49\textwidth]{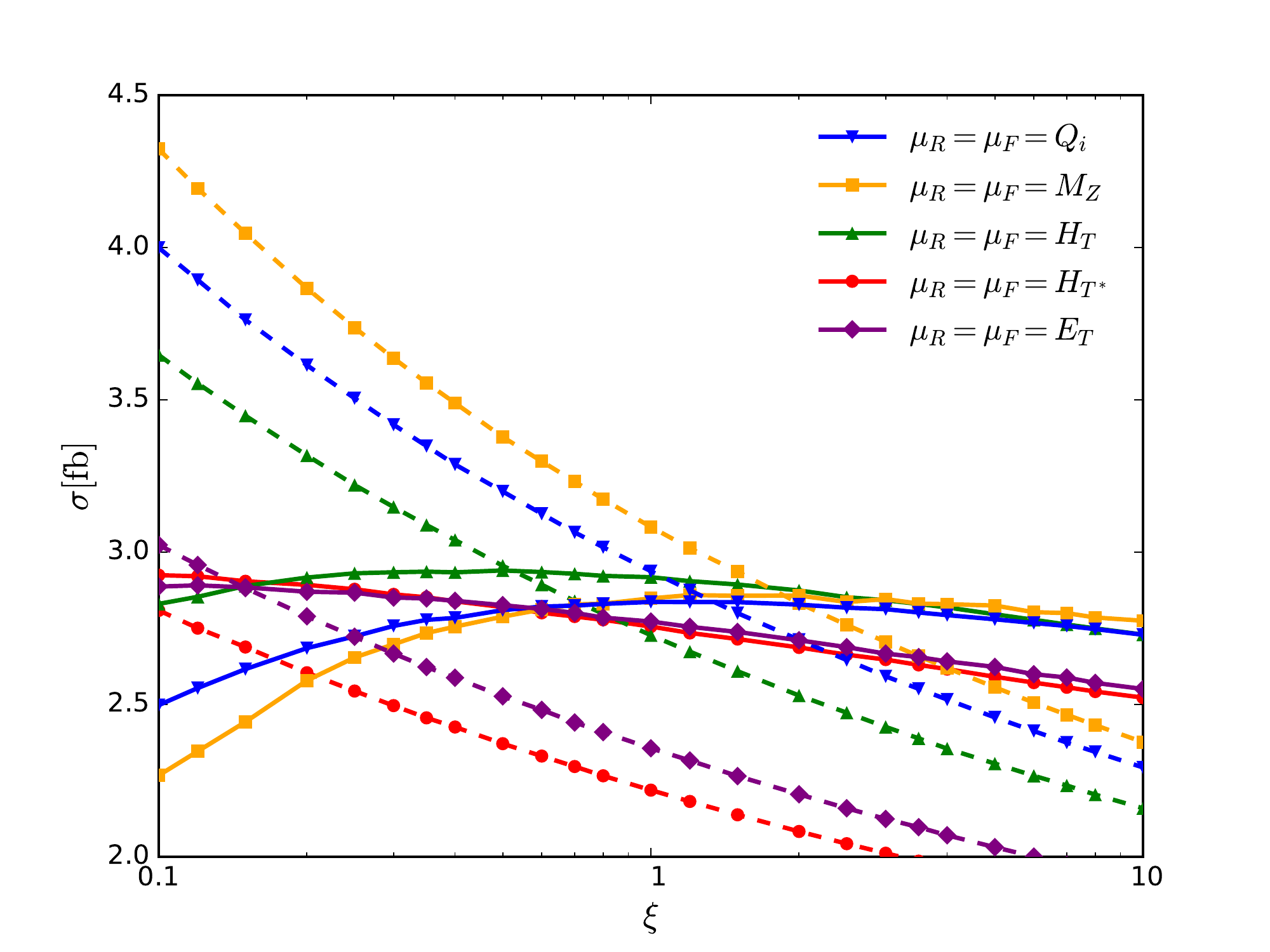}\\
\caption{Left: Dependence of the total cross section on the variation of the factorization and renormalization scale, as well as the combined variation. Results are shown using the central scales $Q_i$ and $M_Z$. Right: Combined variation of the two scales for different choices of the central value.
}
\label{fig:sca0}
\end{figure*}
%
%
%
%
%
%
%
%
%
In the following, we present results for the LHC 
mainly for the \ZAjj channel. Two lepton families are included in the results
presented. Since we apply the same cuts for the first two families,
we compute the process $pp \to e^+ e^- \gamma jj + X,$ and multiply
the result by two.
As EW input parameter, we use the Fermi constant as well as the $Z$
and $W$ mass and derive the remaining EW parameters via tree-level
relations, i.e., we use
%
%
%
\begin{align}
  &G_F = 1.16637 \times 10^{-5} \GeV^{-2},&\nonumber\\         
  &M_W = 80.385 \GeV,                                        
  &M_Z= 91.1876 \GeV, \nonumber \\                           
  &\alpha^{-1}=\alpha^{-1}_{G_F}=132.23422,                     
  & \sin^2 (\theta_W) =0.22289722. \nonumber                 
\label{eq:EWpar}
\end{align}
The widths of the bosons are taken as
\begin{align}
  \Gamma_Z = 2.50773065 \GeV,\quad                           
  \Gamma_W = 2.09666458 \GeV.                                
\end{align}
As default, we use the
PDF4LHC15\_nlo\_100~\cite{PDF4LHC,*Gao:2013bia,*Carrazza:2015aoa} PDF set both
at LO and NLO.
%
%
The jets are defined with the anti-$k_t$ 
algorithm~\cite{Cacciari:2008gp} with radius $R=0.4$ 
and are required to have a transverse momentum $p_{T,j} > 30 \GeV$
%
and rapidity $|y_j|<4.5$. 
The jets are ordered by transverse momenta
and the tagging jets at NLO are defined as the two hardest jets.  
To simulate typical VBF searches and LHC detector capabilities, we use
\begin{align}
p_{T,\ell(\gamma)} &> 20 (30) \GeV , &                       
|y_{\ell(\gamma)}| &< 2.5 , \nonumber\\                      
R_{j\ell} &> 0.4 , &                                       
R_{\ell\gamma} &> 0.4 , \nonumber\\                         
R_{j\gamma} &> 0.4 , &                                     
R_{\ell\ell} &> 0.0, \nonumber \\                          
m_{\ell\ell\gamma} & > 120 \GeV, &                           
m_{\ell\ell} & > 15 \GeV.                                  
\label{eq:zacuts}
\end{align}
The last cut in Eq.~\eqref{eq:zacuts} eliminates the singularity arising
from a virtual photon, $\gamma^* \rightarrow \ell^+ \ell^-$. %
Additionally, we impose the typical VBF cuts on the tagging jets,
\begin{align}
  m_{j_1j_2} & > 600 \GeV,                                
  & |\eta_{j_1}- \eta_{j_2}| & > 4      &    \eta_{j_1} \times \eta_{j_2} < 0.             
\label{eq:zacutsVBF}
\end{align}
%
%
%
%
Furthermore, as a photon isolation cut, following the "tight isolation
accord" presented in Ref.~\cite{Andersen:2014efa}, events are accepted if they satisfy
\bea
\sum_{i\in \text{partons}} p_{T,i}\theta(R-R_{\gamma i}) \le \epsilon p_{T,\gamma}\frac{1-\cos R}{1-\cos\delta_0} \quad \forall R<\delta_0
\label{eq:Frixione cut}
\eea
with $\delta_0 = 0.4$ 
and efficiency, $ \epsilon = 0.05 $.
%
%
%
%


As default factorization and renormalization central scale, we chose the
absolute value of the momentum transfer $Q_i$ from quark line ``i'' 
to the EW process,
\begin{equation}
\label{eq:scale}
\mu_{F} = \mu_{R} = \mu_{0}=Q_i.
\end{equation}

%
%
%
%
%
%
%
%
%
%
%
%
%
%
%
%
%
%
\subsection{Total Cross Section}
%
%
%
%
%
%
%
In the following, we present results for the integrated cross
section at the LHC at a center of mass energy of 13 TeV for different scale choices and
PDF sets.
%
%
%
%
%
%
%
%
%
%
%
In the left panel of Fig.~\ref{fig:sca0}, we vary independently the
factorization and renormalization scale $\mu=\xi \mu_0$ 
in the range $\xi \in (0.1,10)$ around the central scales $\mu_0=M_Z$ (orange
line) and $\mu_0=Q_i$ (blue line). At $ \xi=1 $, we find for
$\mu_0=Q_i$ $ \sigma_{\mathrm{LO}}=2.9378(7) ^{-8\%}_{+9\%} \fb$  and
$\sigma_{\mathrm{NLO}}=2.837(1)^{-0.3\%}_{-1\%} \fb$ with a K-factor,
defined as the ratio of the NLO over the LO predictions, of $K=0.97$.
%
%
Correspondingly, for $\mu_0=M_Z$ we find $
\sigma_{\mathrm{LO}}=3.083(2)^{-8\%}_{+10\%} \fb $ and
$\sigma_{\mathrm{NLO}}= 2.848(4)^{+0.3\%}_{-2\%} \fb$ with a K-factor of $K=0.92$.
The upper(sub)-scripts correspond to the scale uncertainty taken at
$\xi=2 (\xi=0.5)$ being of order $10\%$ at LO and few percent at
NLO. The numbers in parenthesis quote MC statistical errors.
Note that at leading order, we only have factorization scale
uncertainties and that the differences of the LO and NLO predictions
of about $5\%$ and $0.5\%$, correspondingly, at the central scale for
the two different scale choices are contained in the scale
uncertainties. This is, however, not always the case at LO as we can
see in the right panel of Fig.~\ref{fig:sca0} where we have plotted
additionally, setting $\mu_{F} = \mu_{R} = \xi \mu_0$ for simplicity,
the curves for $\mu_0 \in (H_{T},H_{T^{*}},E_T)$, which are often
used in the corresponding QCD $VVjj$ induced processes, and are
defined as:
\begin{align}
  &
H_T=\frac{1}{2}
\left(\sum_{\text{partons}} p_{T,i} + 
\sum_{V_i}\sqrt{p_{T,V_i}^2+m_{V_i}^2}\right),
 \nonumber\\
 &
\HTstar=\frac{1}{2}
\left(\sum_{\text{jets}} p_{T,i} \exp{|y_{i}-y_{12}|} + 
\sum_{V_i}\sqrt{p_{T,V_i}^2+m_{V_i}^2}\right),
\nonumber\\
 & 
E_T=\frac{1}{2} \left[E_{T}(jj) +
  E_{T}(VV)\right], 
\label{eq:scale}
\end{align}%
with $V_i \in (Z,\gamma)$. $m_{V_i}$ denotes the invariant mass of the
corresponding leptons ($m_{V_i}=0$ for on-shell photons) and $y_{12} =
(y_1 + y_2)/2$ the average rapidity of the two hardest (or tagging)
jets, ordered by decreasing transverse momenta. $E_{T}(jj)$ and
$E_{T}(VV)$ stand for the transverse energy of the two tagging jets
and of the $VV$ system, respectively. In the last two scale choices of
Eq.~\eqref{eq:scale}, the first term interpolates between $m_{jj}$ and
$\sum p_{T,jets}$ for large and small $\Delta y_{jj}=|y_1-y_2|$ values, respectively.
%
%
%
%
%
%
%
%
%
%
First, we note that the LO predictions for the different central scale
choices are not covered by the scale uncertainties. At, $\xi=1$, we
found at LO differences of about $40\%$ for $\mu_0=M_Z$ and
$\mu_0=\HTstar$, while the NLO differences are about $5\%$ for
$\mu_0=H_T$ and $\mu_0=\HTstar$. Also, note, that the K-factor varies
from 0.92 for $\mu_0=M_Z$ to 1.24 for $\mu_0=\HTstar$, and greatly
depends on the value of the LO predictions. Even larger differences
can be seen in differential distributions, thus, pointing out the
necessity of using NLO predictions for obtaining robust results.

The order 5\% uncertainty found above for the more extreme scale choices is, 
in fact, a better, but still a low estimate for the error induced by 
missing NNLO corrections. As was found for VBF Higgs production, 
cross sections and various distributions within VBF cuts are lowered 
by another 5 to 10\% by NNLO corrections as
compared to NLO results~\cite{Cacciari:2015jma}. This effect is due to a wider 
energy flow within NNLO quark jets as compared to the NLO approximation: the
narrow $R=0.4$ jets capture less of the jet energy at NNLO and thus have a
harder time passing the $m_{jj}>600$~GeV requirement~\cite{Rauch:2017cfu}.
Since this effect should be universal for all VBF processes, 
we expect another order
5\% reduction of the fiducial cross section at NNLO, which is not accounted
for by the scale considerations.  
\begin{figure}[ht!]
  \centering
  \includegraphics[width=0.49\textwidth]{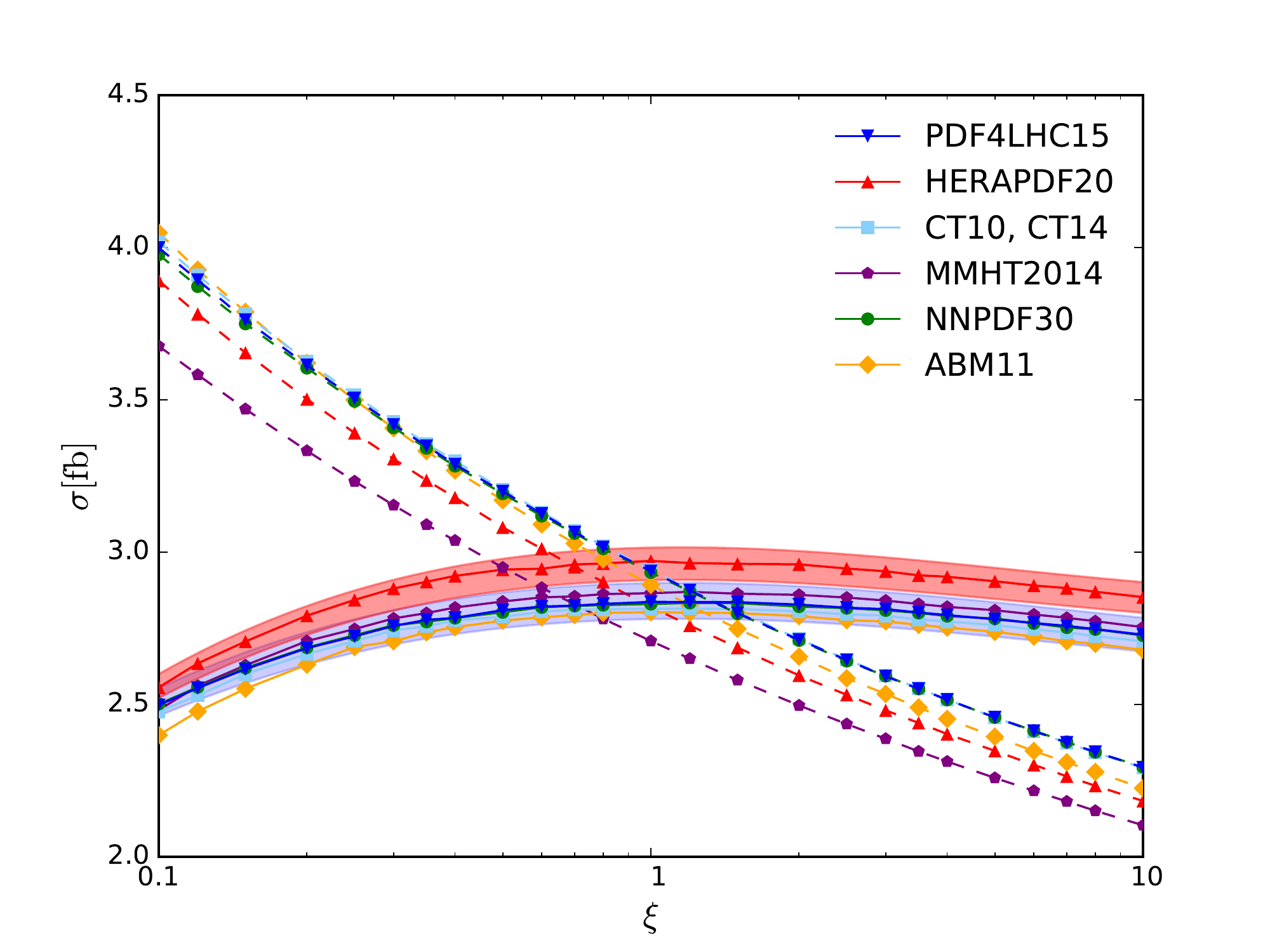}
\caption{Combined variation of the factorization and renormalization scale for different PDF sets.
For the NLO results obtained with the PDF4LHC15 and HERAPDF20 sets, the corresponding PDF uncertainties are shown as bands.
}
\label{fig:sca1}
\end{figure}

\begin{table*}
  \setlength{\tabcolsep}{8pt}
    \caption{Cross section of the EW production process for different PDF sets and associated PDF uncertainties. The factorization and renormalization scale are set equal to $\mu_F=\mu_R=Q_i$.}
   \label{tab:pdf}
  \begin{tabular}{|c|c|c|c|c|c|c|}\hline
  &
  ABM11 &
  CT14 &
  HERAPDF20 &
  MMHT2014 &
  NNPDF30 &
  PDF4LHC15 \\\hline
  $\sigma_{\mathrm{NLO}}      [\fb]$     &
  $2.802(2)^{+1.3\%}_{-0.8\%} $&
  $2.814(2)^{+3.4\%}_{-3.5\%} $&
  $2.972(4)^{+1.8\%}_{-1.7\%} $&
  $2.866(4)^{+2.4\%}_{-2.4\%} $&
  $2.830(6)^{+1.6\%}_{-1.6\%} $&
  $2.837(1)^{+2.1\%}_{-2.1\%} $\\ \hline
  \end{tabular}
 \end{table*}
\setlength{\tabcolsep}{6pt}
In Fig.~\ref{fig:sca1}, we plot the scale variation, for equal
factorization and renormalization scale, i.e.  
$\mu_F=\mu_R=\xi Q_i$, as well as the associated PDF uncertainty at
NLO for the PDF4LHC15\_nlo\_100 and HERAPDF20~\cite{HERAPDF20}~(EIG) sets.
In addition,
we show the scale variation for the
ABM11~\cite{ABM11} (5 flavors),
CT14~\cite{CT14},
MMHT2014~\cite{MMHT14}, 
and
NNPDF30~\cite{NNPDF30} (with $\alpha_s(M_Z) =0.118$)
sets used at
the corresponding order, if available. For NNPDF30 and ABM11, we
use the NLO sets for the LO curves and for the CT14 curve at LO, we use
the LO CT10~\cite{CT10} sets.
At LO, we observe at the central point an $8\%$ maximum difference
between the MMHT2014 and the CT10 predictions, which is of the same
order as the scale uncertainty reported previously, while the error of
the PDFs is around the few percent level and, thus, do not cover the
uncertainty observed.
At NLO, at the central point, a $6\%$ maximum difference between 
HERAPDF20 and AMB11 is found. It is neither covered by the scale
uncertainty, varying only one of the particular scales, nor by the
PDF uncertainties, which 
are  $2\%$ and $1\%$ for these PDFs, respectively. When comparing our
default PDF set (PDF4LHC15\_nlo\_100), with a $2\%$ PDF uncertainty,
with the HERAPDF20 predictions, they almost overlap with a difference
of $4.7\%$. The size of the PDF uncertainties is of the same order in
the whole range $\xi \in (0.1,10)$. 
In Table~\ref{tab:pdf}, one finds the values at NLO obtained with the
different PDFs as well as the associated asymmetric error at  $\xi=1$.

\begin{figure}[ht!]
  \centering
  \includegraphics[width=0.49\textwidth]{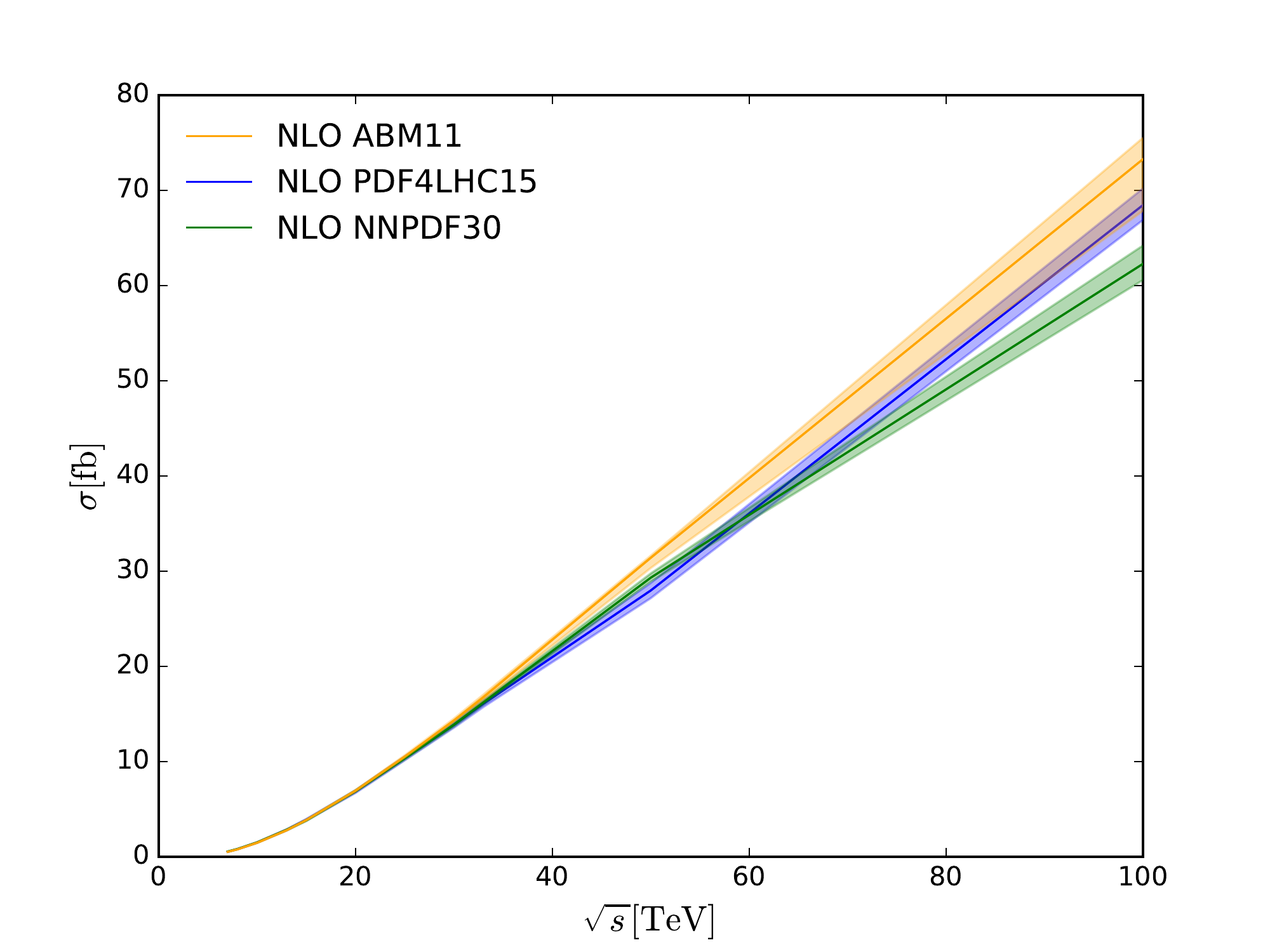}
\caption{Dependence of the total cross section on the center of mass energy using different 
PDF sets. The bands show the PDF and scale uncertainties added in quadrature.
}
\label{fig:ener}
\end{figure}

%
%
%
%

In Fig.~\ref{fig:ener}, we plot the cross sections for different
energies. In addition to our default PDF set, we plot the lines for
the AMB11 and NNPDF30 sets, where larger differences of about 20$\%$
are seen at a center of mass energy of 100 TeV. The bands include the scale and PDF
uncertainties, which are added in quadrature. The combined uncertainty
at 100 TeV rounds the 8$\%$ for the AMB11 set, dominated by the PDF
uncertainty of about 7$\%$, and 3$\%$ for the NNPDF30 and our default
PDF4LHC15\_nlo\_100 set.

\begin{figure}[h!]
  \centering
  \includegraphics[width=0.51\textwidth]{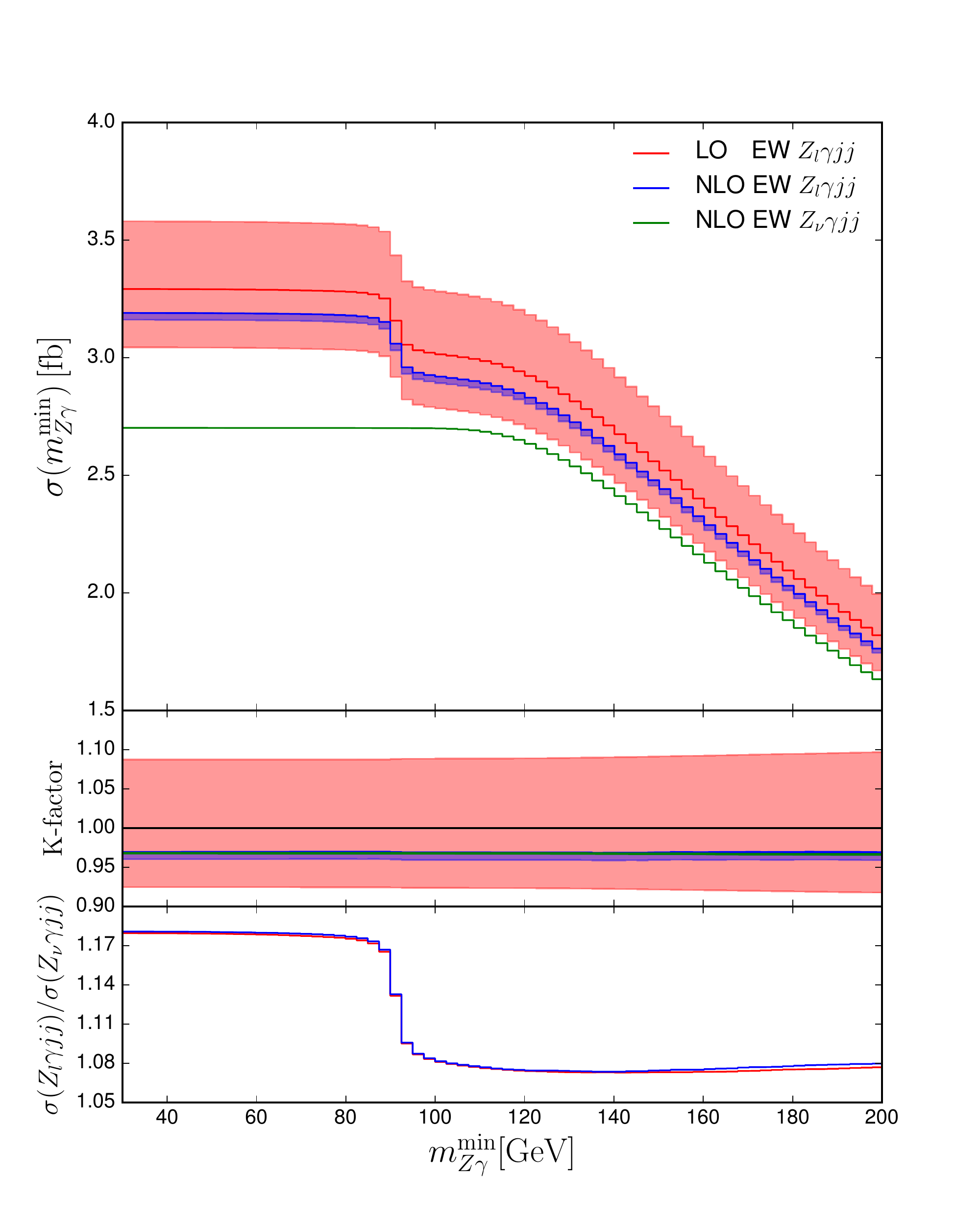}
\caption{The dependence of the total cross section on the minimum required
invariant mass of the EW system is shown in the upper panel. The other panels
show the K-factors and the ratio $\sigma(\ZAjj)/\sigma(\znajj)$.
}
\label{fig:mza}
\end{figure}
In order to remove contributions not relevant for the study of
anomalous coupling effects, in Fig.~\ref{fig:mza}, we plot the
integrated cross section depending on the minimum value required for
the invariant mass, $m_{Z\gamma}$, of the EW system both for the \ZAjj and 
\znajj channels,
neglecting the corresponding cut specified in Eq.~\eqref{eq:zacuts}. 
For leptonic decays of the $Z$-boson, two lepton generations are counted 
while for $Z\to\nu\nu$ only one generation is considered, which gives 
roughly equal cross sections for easier comparison.
When applying cuts, we treat the neutrinos the same as charged leptons, 
such that differences of the two processes can be entirely  associated with 
differences in the amplitude and coupling constants.

For
the \ZAjj channel, we show lines at LO (red) and NLO (blue), 
including
the scale uncertainty, while only the central value at NLO (green) is
shown for the \znajj channel for comparison. In the middle panel, the
K-factors are shown, while in the third panel, the ratios of the \ZAjj and
\znajj channels are plotted.
We observe that imposing a minimum value $m_{Z\gamma} > 90$ \GeV, the
contributions from a photon radiated off the charged leptons (radiative $Z$ decays), which is
absent in the \znajj channel, start to decrease significantly since
configurations of the decay $Z \to \ell^- \ell^+\gamma $ can only be
off-shell. For values of $m_{Z\gamma} > 120$ \GeV, these contributions
are almost completely gone. This is confirmed in the third panel with
a ratio almost constant beyond this value. Similar observations were
found in the QCD induced process in Ref.~\cite{Campanario:2014wga}.

\begin{figure}[h!]
  \centering
\includegraphics[width=0.5\textwidth]{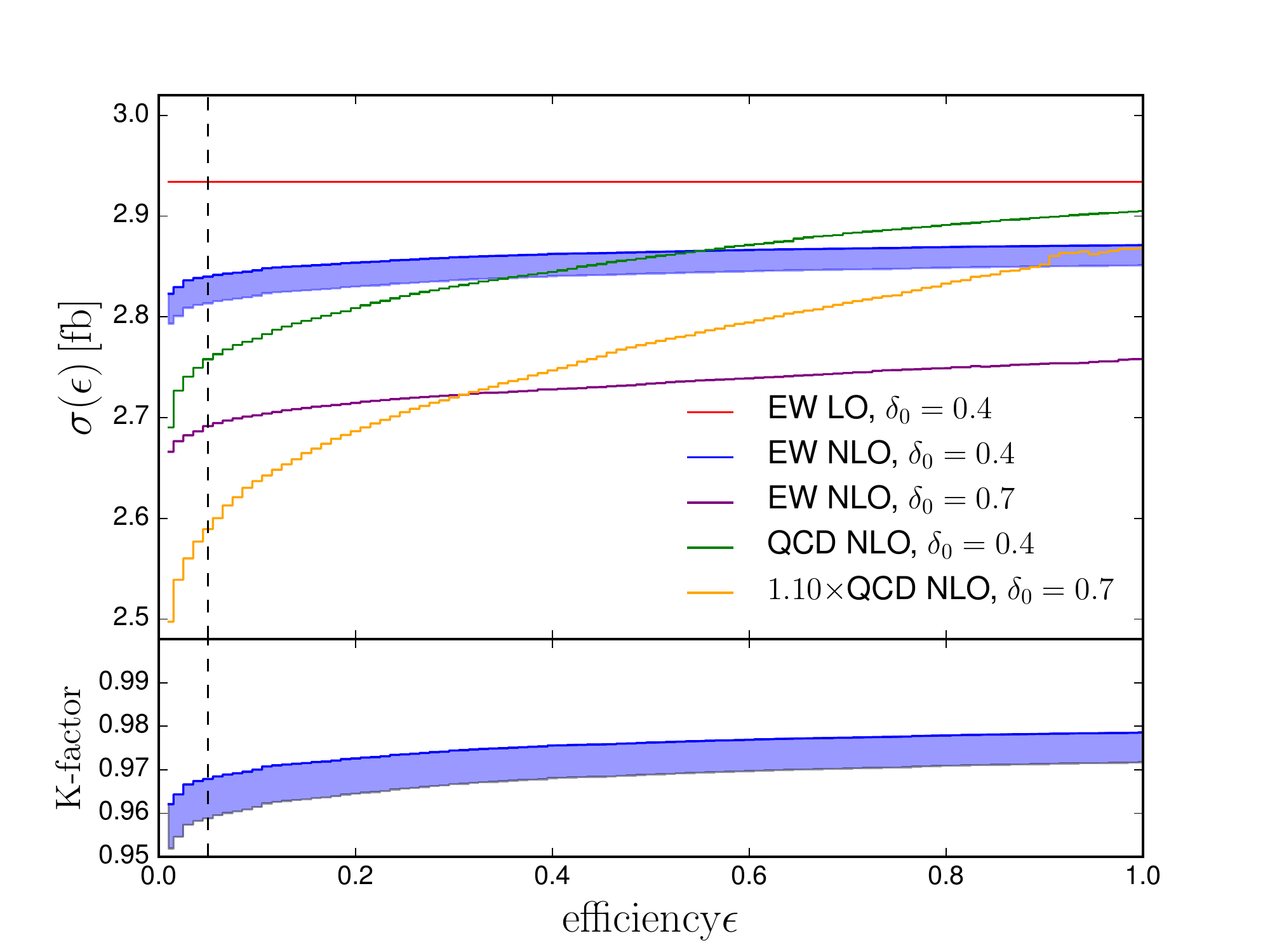}
\caption{Dependence of the total cross section on the photon isolation parameter $\epsilon$
for different sizes of the cone size $\delta_0$.
}
\label{fig:iso}
\end{figure}

%
%
%
%
%
Following the recommendation of the "tight isolation accord" of
Ref.~\cite{Andersen:2014efa}, we set the efficiency to $ \epsilon =
0.05 $. In Fig.~\ref{fig:iso}, we study the dependence of the cross
section on the efficiency parameter plotting the cross section varying
$\epsilon$ in the range $\epsilon \in (0.01,1)$. 
 We observe mild
dependencies of around $3\%(6\%)$ in the whole range shown and of
order $2\%(3\%)$ in $\epsilon \in (0.01,0.05)$ for $\delta_0=0.4
(0.7)$. For comparison, we also show results for the QCD induced
process. In this case, we find larger differences of about
$12\%(23\%)$, with $6\%(13\%)$ differences in the $\epsilon \in
(0.01,0.05)$ range with $\delta_0=0.4 (0.7)$. The milder variation found
in the EW channel is expected due to the characteristic signature of
the VBF processes with two forward jets and the photon produced in the
central region, where little hadronic activity is expected due to the formation
of a rapidity gap in VBF processes.

%
%
%
%
%
%
%
%
%


%
%

Finally, in Table~\ref{tab:ZA} we give the NLO cross sections 
of the EW and QCD
production processes at different collider energies including scale
uncertainties. The number in parenthesis represents the statistical
Monte Carlo integration error, while the upper and the lower numbers
represent the scale uncertainty error at $\xi=2$ and $\xi=0.5$,
respectively. With the given VBF cuts, both
mechanisms are of the same order, in spite of the $(\alpha/\alpha_S)^2$
suppression of the EW channel.

\begin{table}[h!]
    \caption{Cross section of the EW and QCD production process at different center-of-mass energies.
}
    \begin{tabular}{c|cc}
&  EW & QCD \\
  \hline
  8\TeV
  &  $0.808(1)^{-1\%}_{-0.9\%}$ \!fb
  &  $0.735(6)^{-14\%}_{+15\%}$ \!fb \\  
13\TeV
& $2.837(1)^{-0.3\%}_{-1\%}$ \! fb
& $2.764(2)^{-13\%}_{+13\%}$ \! fb \\  
14 \TeV 
& $3.359(6)^{-0.2\%}_{-0.9\%}$ \! fb %
& $3.31(2)^{-12\%}_{+13\%}  $ \!  fb \\%
  \end{tabular}
    \label{tab:ZA}
\end{table}

\begin{figure*}[ht!]
  \centering
\includegraphics[width=0.5\textwidth]{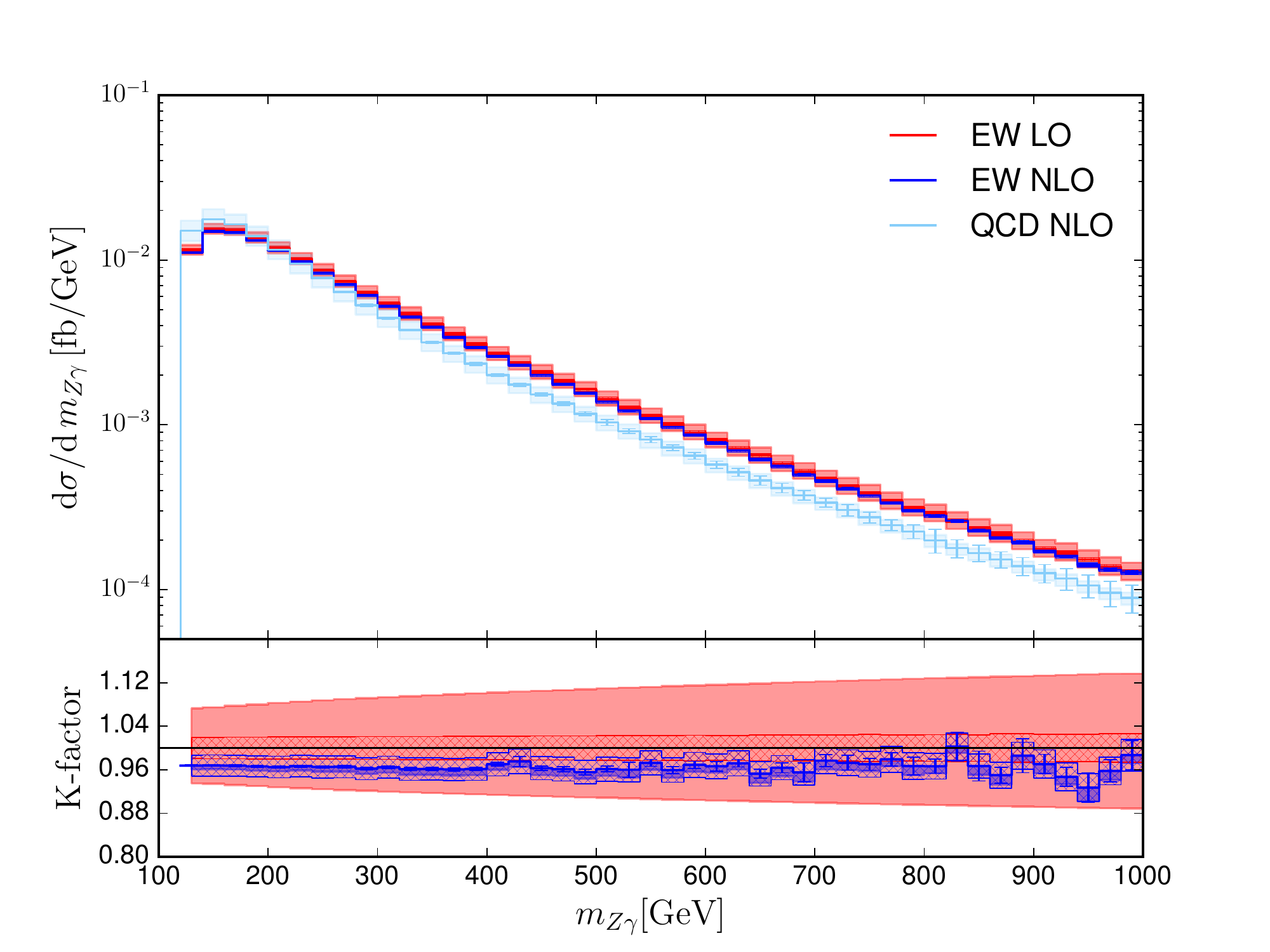}\hfill
\includegraphics[width=0.5\textwidth]{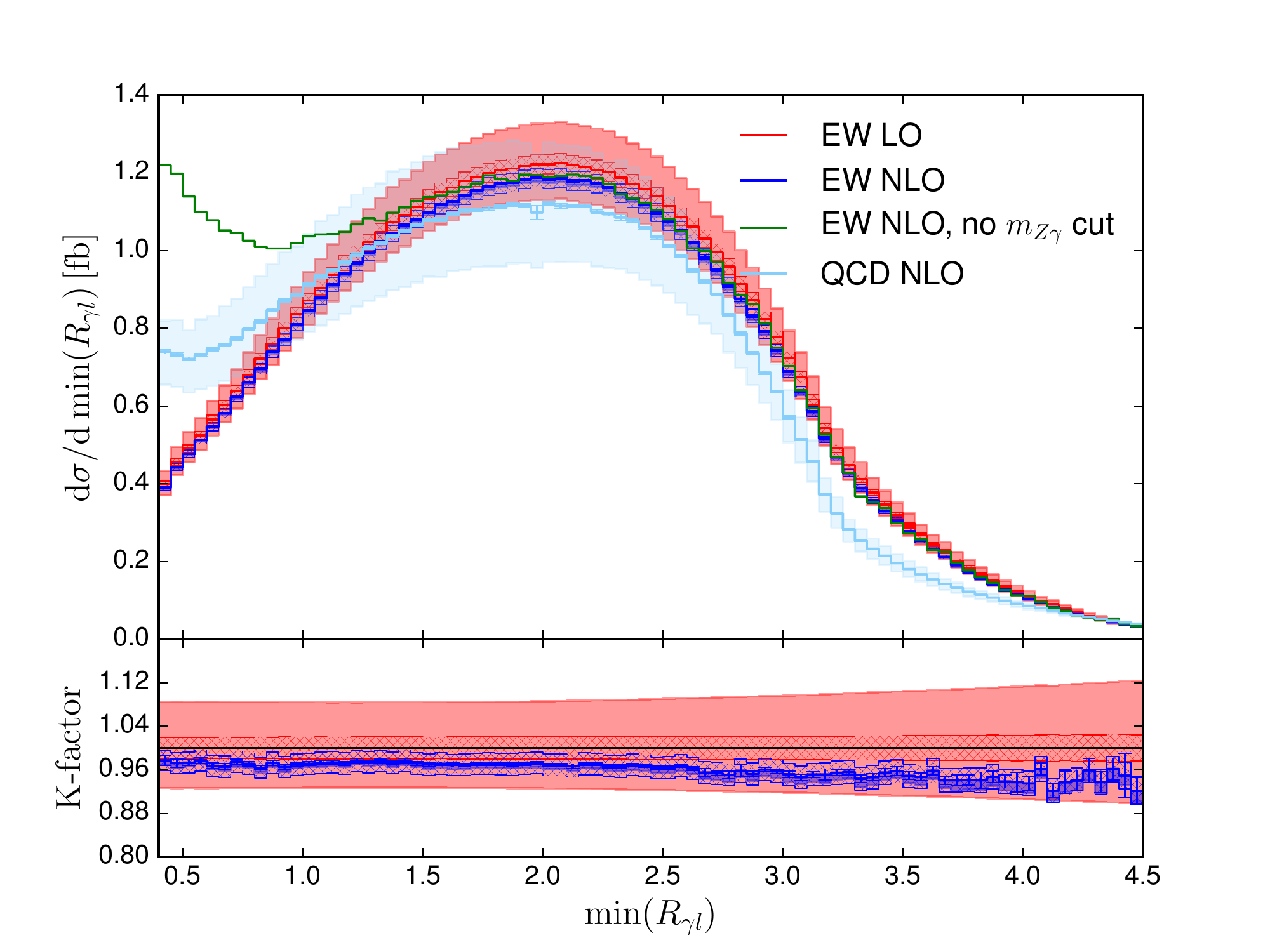}
\includegraphics[width=0.5\textwidth]{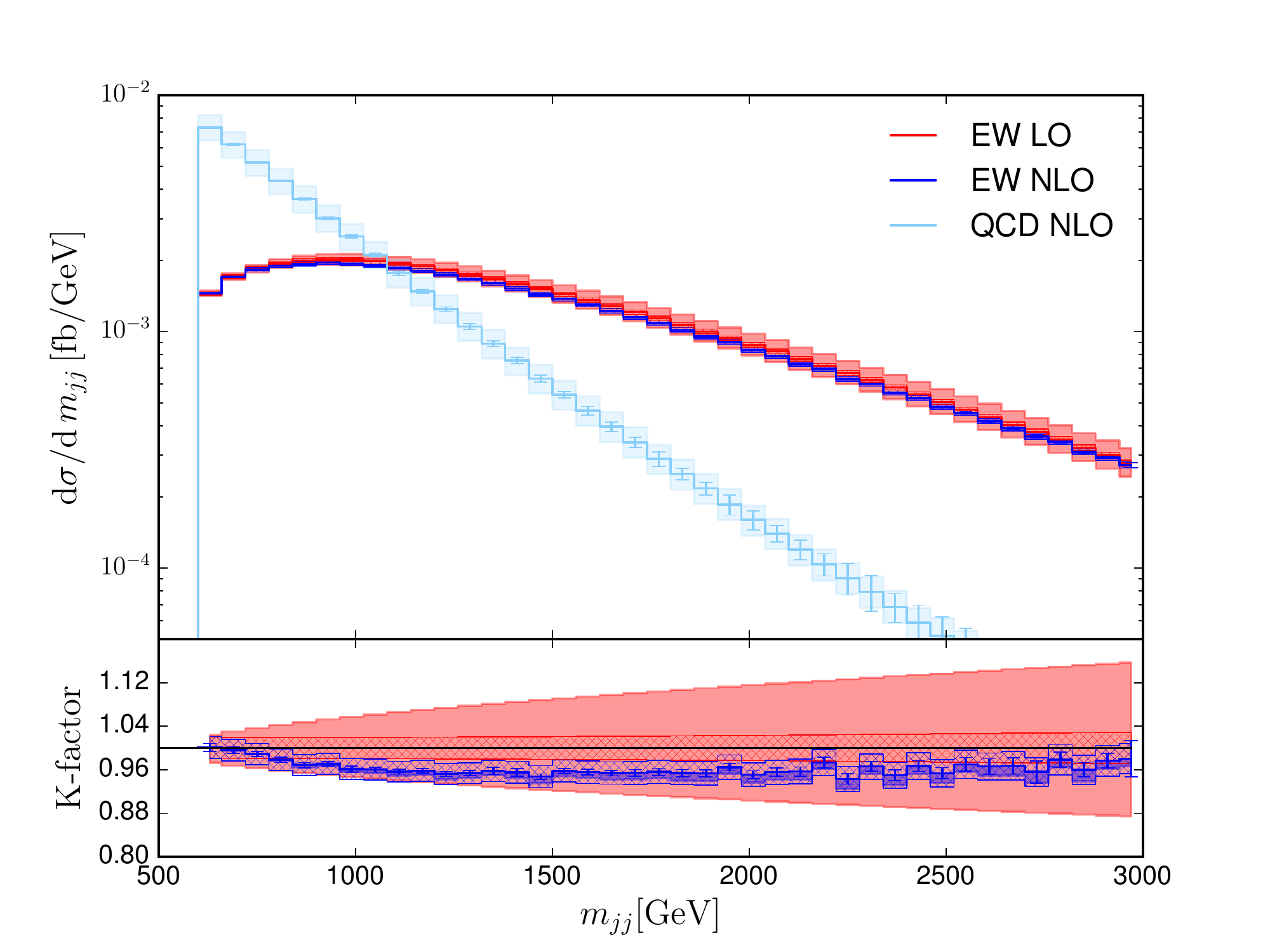}\hfill
\includegraphics[width=0.5\textwidth]{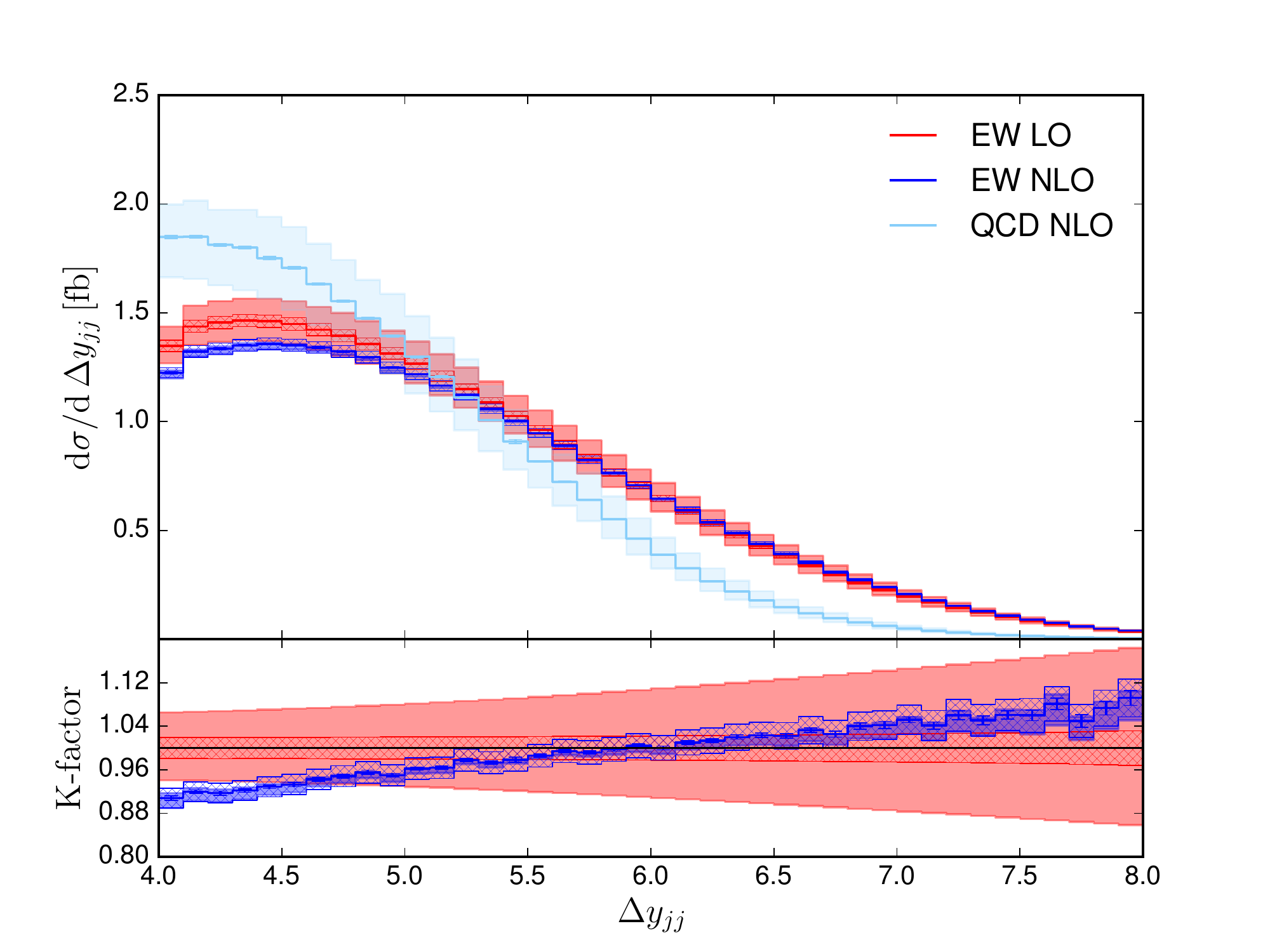}
\caption{Differential cross sections of the EW and QCD induced process,
showing the dependence on $m_{Z\gamma}$, $\mathrm{min}(R_{\gamma l})$, $m_{jj}$, and $\Delta y_{jj}$.
For the $\mathrm{min}(R_{\gamma l})$ distributions, we also show the NLO EW cross section without 
applying a cut on $m_{Z\gamma}$. 
Solid bands result from scale variation by a factor of two around the central value. For the EW process, the uncertainties associated with the PDF
are shown as hatched bands.
}
\label{fig:diff0}
\end{figure*}

%
%
%
%
%
%
%
%
%

\subsection{Differential Distributions}

In the following, we show results for differential distributions at
$\sqrt{s}=13\TeV$ for the QCD and EW induced \ZAjj processes using our
default settings described in section~\ref{sec:calc}. In the top
panels, we show the EW LO (red) and NLO (dark-blue) curves,
including the scale uncertainties. In light-blue, we show the QCD
induced process at NLO, including its scale uncertainties. In the
bottom panels, we show the corresponding EW K-factor as well as the
scale uncertainty band compared to the LO result at the central scale.
PDF uncertainties of the EW process are shown as hatched bands.
In Fig.~\ref{fig:diff0}, we show in the upper row the differential
distribution of the invariant mass of the electroweak system
(left)  and the minimum
R-separation between the photon and one of the leptons (right).
In the lower row, we show jet observables for the tagging jets, the
dijet invariant mass (left) and the rapidity separation (right).
Given the appropriate scale choice, $Q_i$, we observed modest K-factors, close
to one, in the whole spectrum, with larger variation in the rapidity
separation plot, ranging from 0.90-1.10, and a drastic reduction of the
scale uncertainties.
In the top-right plot, we show in addition the curve (green) without
the $m_{ZA}>120 \GeV$ cut. As expected, the cut only reduces events with 
photons emitted close to the charged leptons.
In the bottom-left plot, one can observe clearly the distinct
behaviour of the invariant mass distribution of the tagging jets for the
EW vs QCD channels, with a steeper fall-off of the
cross section for the QCD induced process.
\begin{figure*}[ht!]
  \centering
\includegraphics[width=0.5\textwidth]{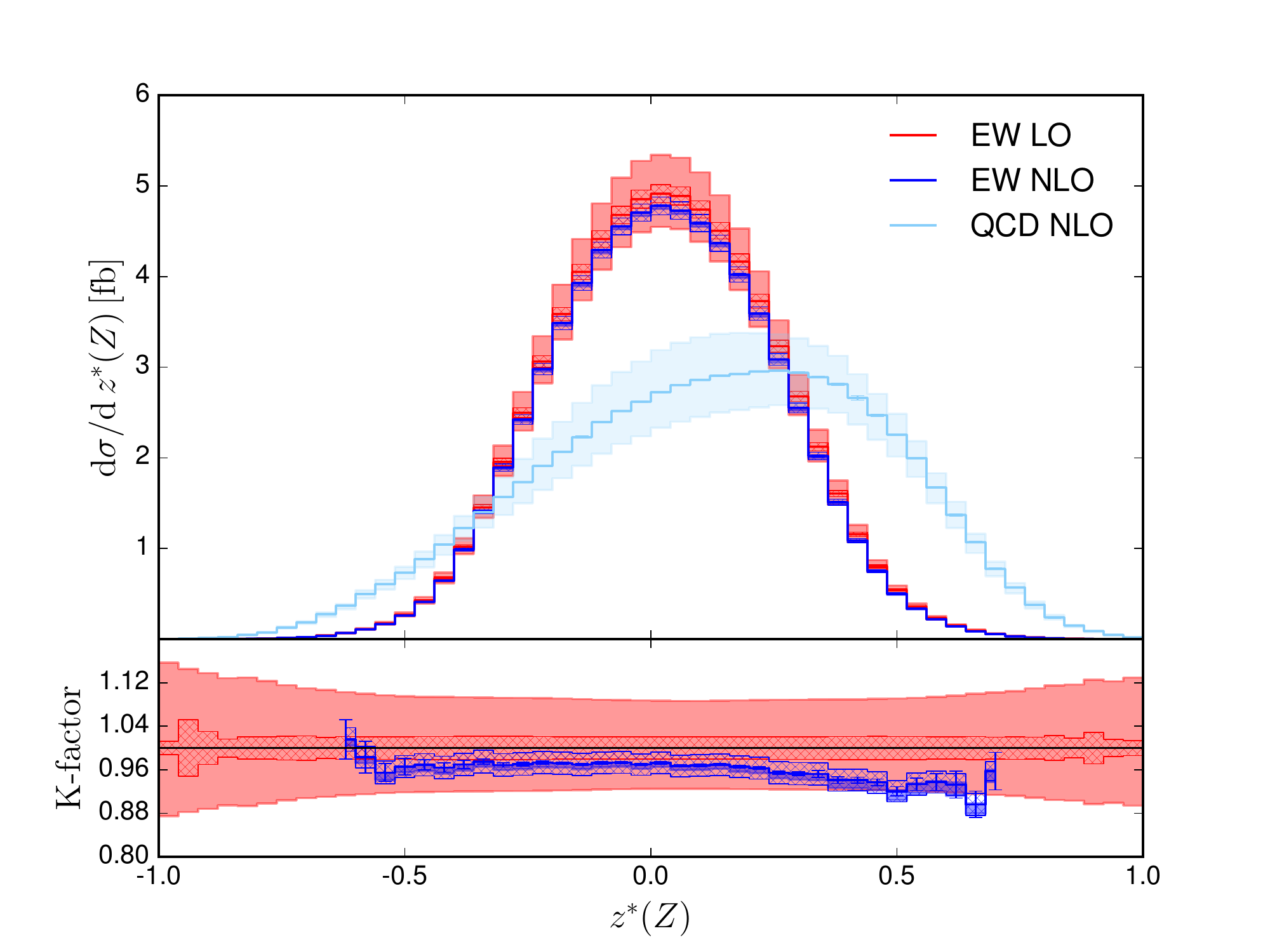}\hfill
\includegraphics[width=0.5\textwidth]{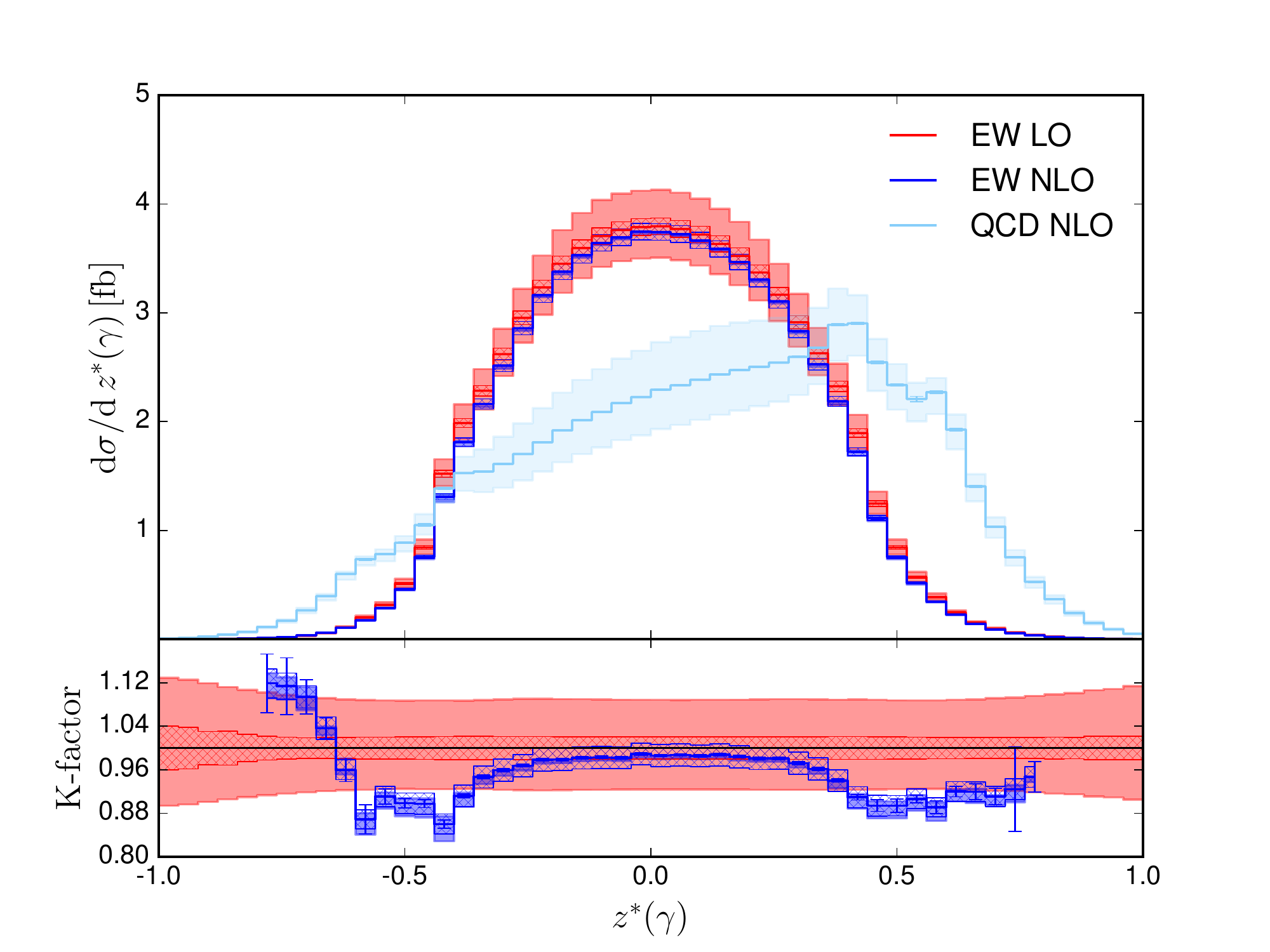}
\caption{Differential cross sections of the EW and QCD induced process,
showing the dependence on $z^*$ defined in Eq.~\eqref{eq:zstar}.
The bands show uncertainties associated with the scale and PDF set as described in Fig.~\ref{fig:diff0}.
}
\label{fig:diff1}
\end{figure*}

In Fig.~\ref{fig:diff1}, we show the normalized centralized rapidity
distribution of the reconstructed Z boson system (left) and the photon
(right) with respect to the tagging jests,
\begin{align}
z^*(V)=\frac{y_V - \frac{1}{2} (y_1+y_2)}{y_1-y_2}.
\label{eq:zstar}
\end{align} 
Whereas for the EW process, the electroweak particles are nearly exclusively produced in the central region between the two tagging jets 
located at $z^*=\pm 1/2$, they are produced in a broader rapidity range for the QCD process. 
Note that the distributions are not symmetric because the jets are 
$p_T$-ordered.  In particular for the QCD process, larger contributions can be found in the vicinity of the hardest jet. The particular shape of the QCD distributions can be explained by kinematic configurations, where the hardest jet recoils against the EW system, and a second jet, possibly stemming from gluon radiation, is produced at large separation to fulfill the VBF cuts.

%
%
%
%
%
%
%
%
%

\subsection{Anomalous Couplings}
In our implementation of the \ZAjj production cross section, we allow for modified gauge couplings in the framework of an effective Lagrangian 
\begin{align}
  \mathcal L_{EFT} = \mathcal L_{SM} + \sum_{d>4}\sum_i \frac{f_i}{\Lambda^{d-4}}\mathcal{O}_i^{(d)},
  \label{}
\end{align}
where the operators of dimension 6 and 8 have first been defined in Refs.~\cite{Hagiwara:1993qt,Hagiwara:1993ck,Eboli:2006wa}. Due to minor differences in the definition of the field strength tensors in VBFNLO, our conventions for the dimension 8 operators differ from the ones given in Ref.~\cite{Eboli:2006wa}. The exact definition, as well as the corresponding conversion rules can be found in Ref.~\cite{Baglio:2014uba}.
While all operators given in Refs.~\cite{Hagiwara:1993qt,Hagiwara:1993ck,Eboli:2006wa} affect the \ZAjj production cross section, the operators
\begin{align}
  \mathcal O_{T,8} &= \widehat{B}_{\mu\nu}\widehat{B}^{\mu\nu} \widehat{B}_{\alpha\beta}\widehat{B}^{\alpha\beta} \quad \text{and}\\
  \mathcal O_{T,9} &= \widehat{B}_{\alpha\mu}\widehat{B}^{\mu\beta} \widehat{B}_{\beta\nu}\widehat{B}^{\nu\alpha}
\end{align}
with
\begin{align}
  \widehat{B}_{\mu\nu} = i \frac{g'}{2} B_{\mu\nu}^a 
\end{align}
are of particular interest for \ZAjj production since they only involve neutral gauge bosons. Hence, they can first be constrained in vector boson scattering $pp\rightarrow VVjj$ or triboson production $pp\rightarrow VVV$ of neutral gauge bosons ($V\in(Z,A)$). Current experimental constraints on these operators can be found in Refs.~\cite{Aad:2016sau,Khachatryan:2017jub}.

Including anomalous gauge couplings, the amplitude rises 
as $\mathcal M(s)\propto s^2$ for large invariant masses $s=m^2_{Z\gamma}$ of the underlying vector boson scattering process, leading to unitarity violation for large invariant masses.
Unitarity of the scattering amplitude can be restored by multiplying the amplitude with a form factor of the form
\begin{align}
  \mathcal F(s) = \left( 1+\frac{s}{\Lambda^2_{FF}} \right)^{-2}.
	\label{eq:FF}
\end{align}
A different approach to unitarize the amplitude, via $K$-matrix unitarization, 
has been explored in Ref.~\cite{Kilian:2014zja}, 
leading to a modification of the normalized eigen-amplitudes $a_{IJ}$ 
according to 
\begin{align}
  a_{IJ}\rightarrow \frac{a_{IJ}}{1-i a_{IJ}}.
  \label{eq:kmatrixUnitarization}
\end{align}
In the large $s$ limit, $K$-matrix unitarization leads to a behavior similar to applying a modified, complex form factor~\cite{Loeschner},
\begin{align}
  \mathcal F^c(s) = \left( 1-i \frac{s^2}{{\Lambda^{c}_{FF}}^4} \right)^{-1},
	\label{eq:FFc}
\end{align}
and in the following we compare this complex form factor with the conventional form factor defined in Eq.~\eqref{eq:FF}.
The form factor scales $\Lambda_{FF}$ and $\Lambda_{FF}^c$ are set according to the unitarity constraint, such that the helicity combination with the largest contribution to the zeroth partial wave fulfills the unitarity condition for all vector boson scatterings $VV\rightarrow VA$ and $WW\rightarrow VA$ ($V\in(Z,A)$). 

\begin{figure}[ht!]
  \centering
\includegraphics[width=0.45\textwidth]{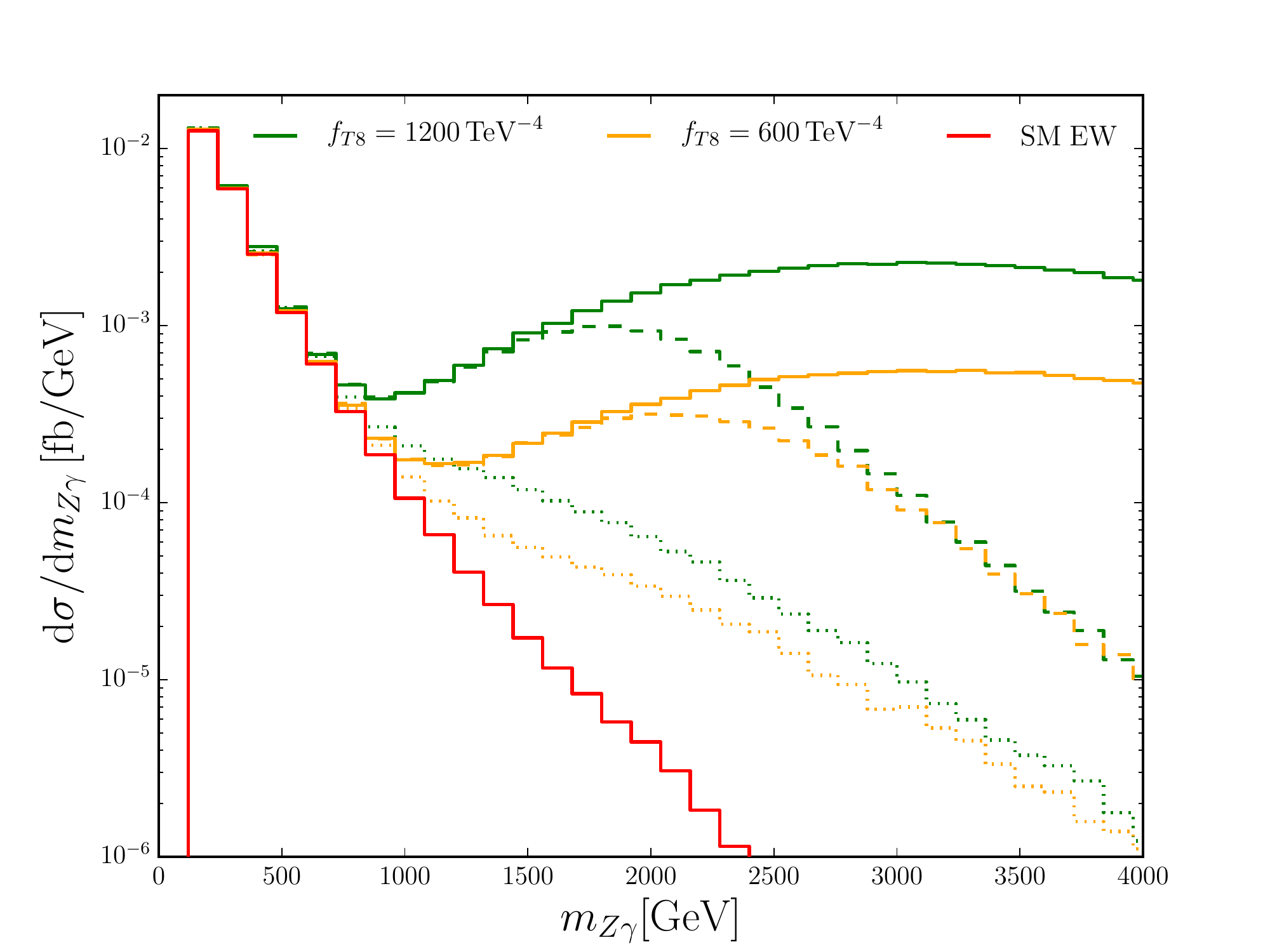}
\caption{Cross section of \ZAjj production in the SM and for different values  of $f_{T8}$. 
The line styles are as given in \fig{fig:AC1}.
  }
\label{fig:ACmVV}
\end{figure}

In \fig{fig:ACmVV}, we show the dependence of the \ZAjj cross section on the invariant mass $m_{Z\gamma}$ of the electroweak system for different values of $f_{T8}$.  It can be seen that well below the form factor scale, the results using the complex form factor (dashed lines) closely follows the results where no unitarization is applied. Only close to and above the form factor scale, the modified form factor leads to a significant suppression, removing the unitarity violating tail of the distribution. The different values of $f_{T8}$ influence the shape of the distribution well below the form factor scale, but for very high invariant masses they lead to identical results, given by the unitarity condition. 
In contrast, the conventional form factor (dotted lines) also leads to a significant reduction of the cross section well below the form factor scale, reducing the effect of the anomalous coupling in a larger region of the phase space. 
\begin{figure*}[ht!]
  \centering
\includegraphics[width=0.9\textwidth]{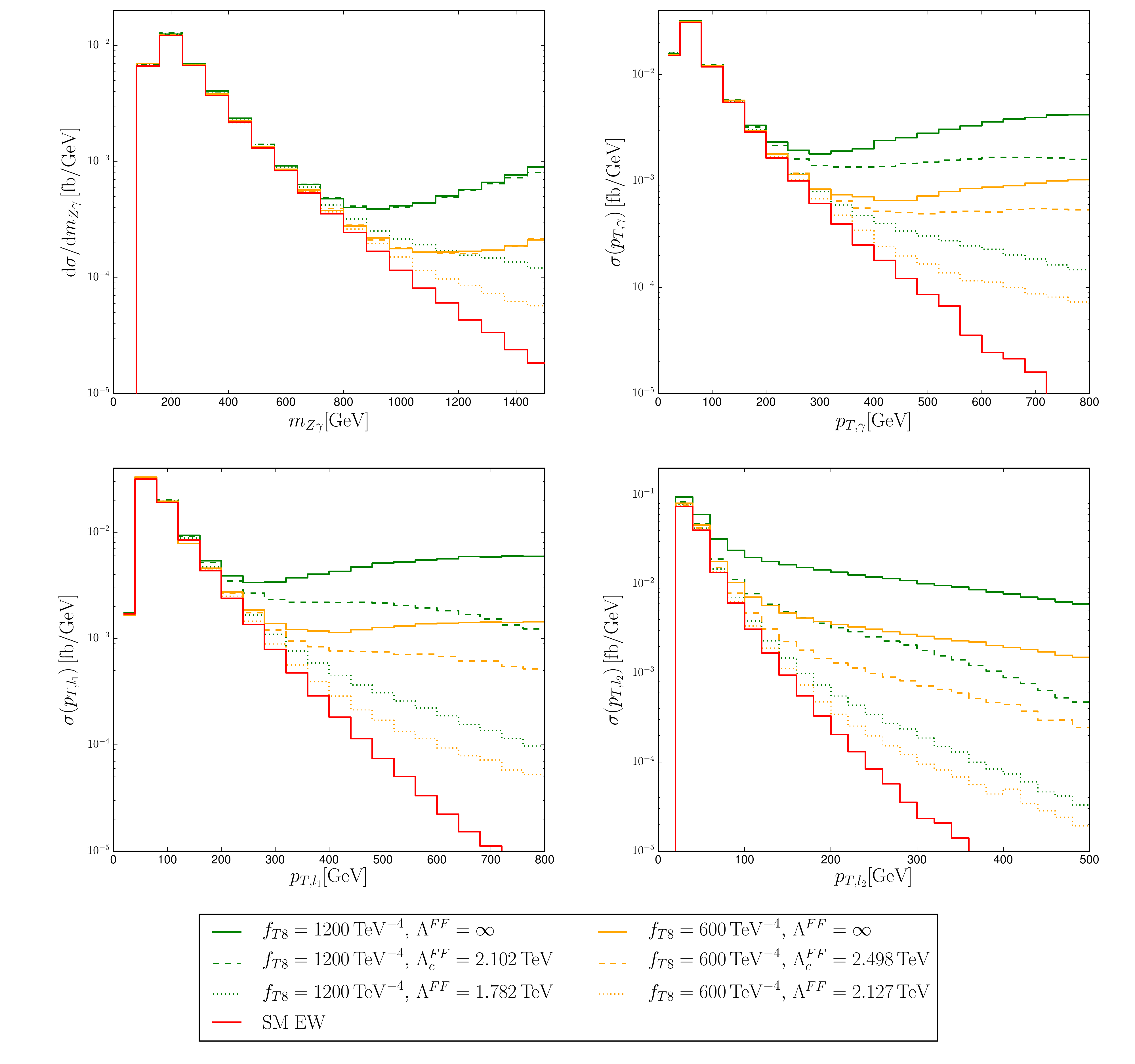}
\caption{Cross section of \ZAjj production in the SM and including various anomalous couplings in dependence on $m_{ZA}$ and transverse momentum of the final state particles.}
\label{fig:AC1}
\end{figure*}

\begin{figure}[ht!]
  \centering
\includegraphics[width=0.45\textwidth]{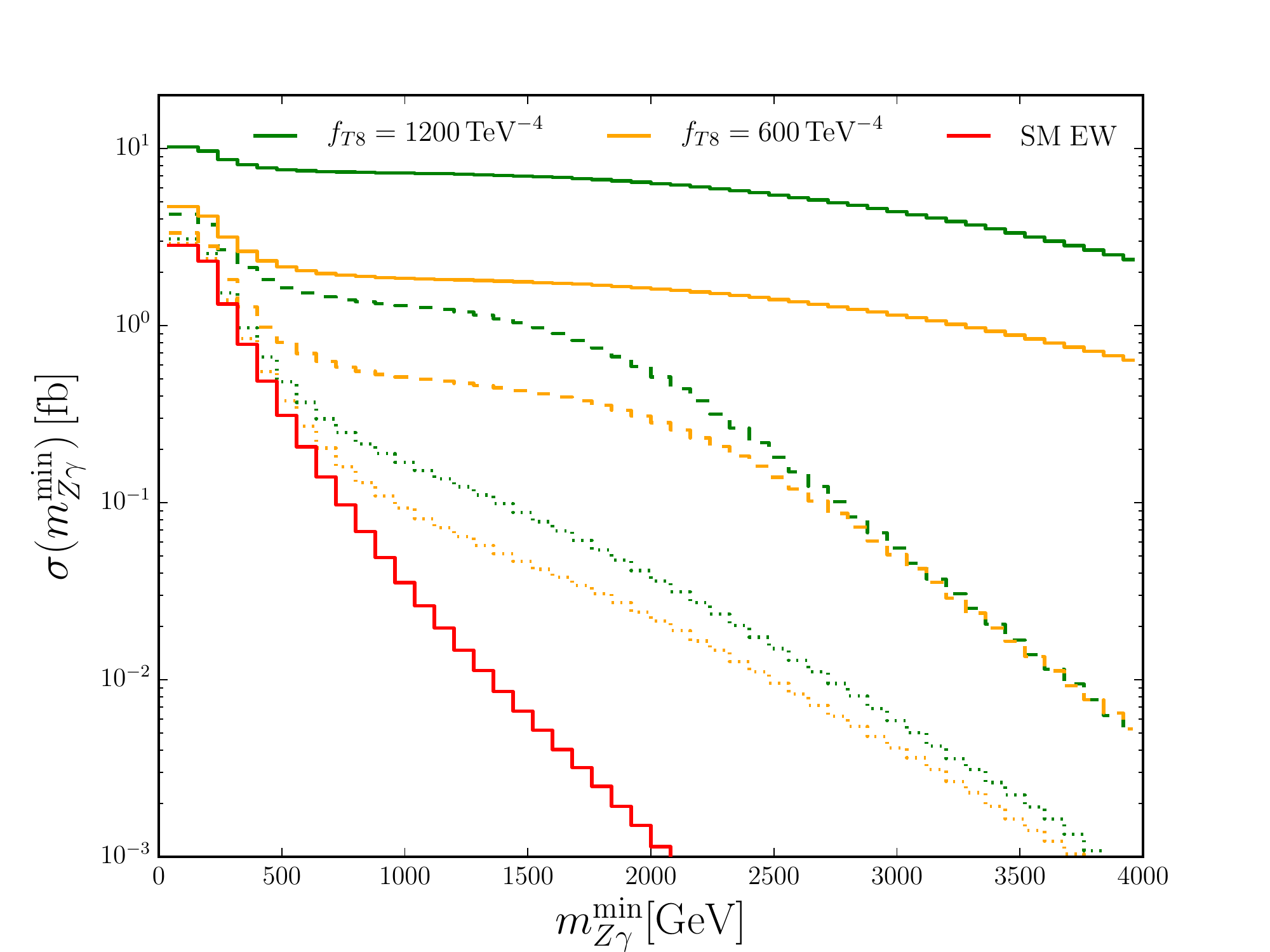}
\caption{Dependence of the \ZAjj total cross section on an additional cut $m_{Z\gamma}^{\mathrm{min}}$ according to Eq.~\eqref{eq:AC_mZAmin}. The line styles are as given in \fig{fig:AC1}.}
\label{fig:ACmVV_int}
\end{figure}
While \fig{fig:ACmVV} illustrates the differences of the two form factors over a broad range of $m_{Z\gamma}$, it is clear that the phase-space region above the form factor scale is determined by the unitarization procedure and shouldn't be used to constrain anomalous couplings. In \fig{fig:AC1} we focus on smaller invariant masses and, in addition, we show the cross section differential in the final state transverse momenta. Similar to the $m_{Z\gamma}$ distribution, we observe that the conventional form factor (dotted lines) leads to a large suppression, whereas  its complex version (dashed lines) leads to results much closer to the results without unitarization. However, also for the complex form factor, the deviations from the result without unitarization start already at small transverse momentum. In particular for the transverse momentum of the softer lepton, we obtain a large suppression already for very small values of $p_{T,l_2}$.

Since anomalous gauge couplings lead to an increased cross section in the tails of the distributions, experimental limits are often obtained from a comparison of the observed event count with the cross section 
\begin{align}
  \sigma(m_{Z\gamma}^{\mathrm{min}}) = \int_{m_{Z\gamma}^{\mathrm{min}}}^\infty \mathrm dm_{Z\gamma} \frac{\mathrm d\sigma(m_{Z\gamma})}{\mathrm d m_{Z\gamma}},
\label{eq:AC_mZAmin}
\end{align}
where a lower cut on the invariant mass of the electroweak system is applied.  We therefore show the dependence of the cross section on this cut in \fig{fig:ACmVV_int}, where 
it can be seen that a large fraction of the cross section results from contributions with invariant masses above the form factor scales. We want to point out that theoretical predictions in this phase space region highly depend on the unitarization procedure. A preferred procedure to compare experimental results with theoretical predictions should therefore be a comparison based on differential distributions, restricted to invariant masses of the electroweak system well below the form factor scale.




\section{Conclusions}
\label{sec:concl}
In this article, we have reported results at NLO QCD for VBF $Z\gamma$
production, including the leptonic decay of the $Z$ boson with all
off-shell effects and spin correlations taken into account.
While, at LO, the results greatly depend on the scale choice, with up
to 40$\%$ differences at the central value, the NLO results reduce
considerably the scale uncertainties, to the few percent level.
PDF uncertainties of individual sets have been studied yielding errors
of a few percent, which are propagated quite homogeneously over the available
phase space. However, the central values of the predictions for the
different sets differ by up to 6$\%$, which is not covered by the combined
uncertainty associated to the pdf sets and scale variation.
Furthermore, we have presented results for anomalous couplings and we 
pointed out the
necessity to constrain anomalous couplings with data 
well below the form factor scale only.

\section{Acknowledgments}
FC acknowledges discussions with J.~Bernabeu and financial support by
the Spanish Government and ERDF funds from the European Commission
(Grants No. RYC-2014-16061, FPA2014-53631-C2-1-P , FPA2014-57816-P,
and SEV-2014-0398). The work of DZ was supported by "BMBF
Verbundforschung Teilchenphysik" under grant number 05H15VKCCA.



\bibliographystyle{h-physrev}
\bibliography{QCDVVjj}

\end{document}